\documentclass[reprint,amsmath,amssymb,aps,prb]{revtex4-2}
\usepackage{graphicx,dcolumn,bm,tikz,times,physics,xcolor}
\usepackage[pdftex,colorlinks,linkcolor={blue!75!black},citecolor={blue!75!black},urlcolor={blue!75!black}]{hyperref}
\usepackage{amsfonts,amssymb,amsmath}

\newcommand{\rmi}{\ensuremath{\mathrm{i}}}
\newcommand{\rme}{\ensuremath{\mathrm{e}}}

\newcommand{\hH}{\ensuremath{\hat{H}}}
\newcommand{\hc}{\ensuremath{\hat{c}}}

\newcommand{\cT}{\mathcal{T}}
\newcommand{\cN}{\mathcal{N}}
\newcommand{\cH}{\mathcal{H}}
\newcommand{\cU}{\mathcal{U}}
\newcommand{\cV}{\mathcal{V}}
\newcommand{\9}{\left(}
\newcommand{\0}{\right)}
\newcommand{\zf}{\left[}
\newcommand{\yf}{\right]}

\newcommand{\bfu}{\ensuremath{\mathbf{u}}}

\begin{document}

\title{Loss-induced universal one-way transport in periodically driven systems}

\author{Chang Shu}
\author{Kai Zhang}
\email{phykai@umich.edu}
\author{Kai Sun}
\email{sunkai@umich.edu}
\affiliation{Department of Physics, University of Michigan, Ann Arbor, Michigan 48109, United States}
% \date{\today}

\begin{abstract}
In this paper, we show that a 
periodically driven Aubry-Andr\'e-Harper model with imbalanced on-site gain/loss supports universal one-way transport that is immune to impurities and independent of initial excitations. 
We reveal the underlying mechanism that the periodic driving gives rise to the non-Hermitian skin effect in the effective Floquet Hamiltonian, thereby causing universal non-reciprocal transport.
Additionally, we probe the time-average decay rate of the propagator under long-time bulk dynamics as a signature of the Floquet emergent non-Hermitian skin effect. 
Our results provide a feasible and controllable way to realize universal one-way transport that is easily accessible to experiments. 
\end{abstract}

\maketitle
\section{Introduction}
Since the discovery of chiral edge currents in two-dimensional (2D) quantum hall systems~\cite{von1986quantized,klitzing1980new,haldane1988model} and topological insulators~\cite{bernevig2006quantum,hasan2010colloquium}, one-way transport immune to disorders and defects has drawn extensive research interest due to its wide potential applications~\cite{enhanced2104xu,wang2013quantum,zhang2012chiral,Torres2016crafting,Berdakin2021}.
In conventional one-dimensional (1D) band systems, however, backscattering-free one-way transport is prohibited by the fermion doubling theorem~\cite{nielsen1981no,nielsen1981absence1,nielsen1981absence2}, which states that the left and right propagating modes in a 1D band system must appear in pairs. 
Various strategies have been attempted to bypass this constraint. 
So far, realizing one-way transport in Hermitian band systems is still experimentally challenging due to the requirement of engineering long-range interactions~\cite{longhi2016robust,nason1985lattice,wang2022validity,KAPLAN1992342}. 

Non-Hermitian band systems~\cite{ashida2020non} can evade the fermion doubling owing to the non-Hermitian skin effect~\cite{Yao2018,kunst2018biorthogonal,Torres2018,ChingHua2019,Murakami2019,zhang2020correspondence,RobertJan2020,yang2020aGBZ,Okuma2020}, where the eigenstates of system-size order concentrate at edges of the 1D open chain. 
Under a given excitation frequency, the rightward and leftward propagating waves exhibit different lifetimes in the 1D bulk, leading to the unidirectional transport under long-time dynamics~\cite{Ashvin2019PRL,Kai2022arXiv,Longhi2015oneway}. 
One possible pathway to achieve such one-way propagation is to introduce non-reciprocal hoppings between neighboring sites. 
However, the implementation of non-reciprocity demands sophisticated experimental design~\cite{hatano1996model,funnellight2020,generatewinding2021,XuePeng2020,Ghatak2020}. 
A more natural and flexible way utilizes imbalanced on-site dissipation, as they are inevitable in most classical wave systems and can be easily manipulated in mechanical metamaterials~\cite{bertoldi2017flexible,Ghatak2020,Jayson2022PRA}, photonic and acoustic crystals~\cite{Regensburger2012,FengLiang2017,cheng2020experimental,zhou2018observation,kruss2022}, and cold-atom chains~\cite{diehl2008quantum}. 
However, one needs to either break the time-reversal symmetry~\cite{SatoPRX2019,YiYang2020PRL,Kawabata2020PRB} of the original Hamiltonian or to control next-nearest-neighboring hoppings~\cite{lee2016anomalous,li2023lossinduced, sun2024}. 
Both are challenging tasks for the systems where most non-Hermitian effects have been observed~\cite{XuePeng2020,Ghatak2020,generatewinding2021,LQPRL2022}.

Floquet systems~\cite{RudnerNRP2020,rudner2013PRX,Rechtsman2013,Demler2010PRB,WeitenbergNP} offer another platform to achieve one-way transport. 
The periodicity of the Floquet quasienergies allows an energy band to be one-way tilted, a seemingly promising feature to realize one-way transport~\cite{PeterZoller2017,Torres2017laser,Hockendorf2020,titum2016PRX,PriviteraPRL2018}.
Nevertheless, a no-go theorem~\cite{sun2018chiralmagnetic} still holds for the full quasienergy spectrum, ensuring the existence of counter-propagating modes at the stroboscopic level. 
Consequently, one-way transport in such systems is not universally immune to all types of impurities~\cite{PeterZoller2017,Hockendorf2020}, and can only be achieved if one carefully avoids triggering the counter-propagating modes. 
Is there a general approach to achieve robust and experiment-friendly one-way transport, eliminating counter-propagating modes altogether?

\begin{figure*}[t]
    \centering
    \includegraphics[width=1\linewidth]{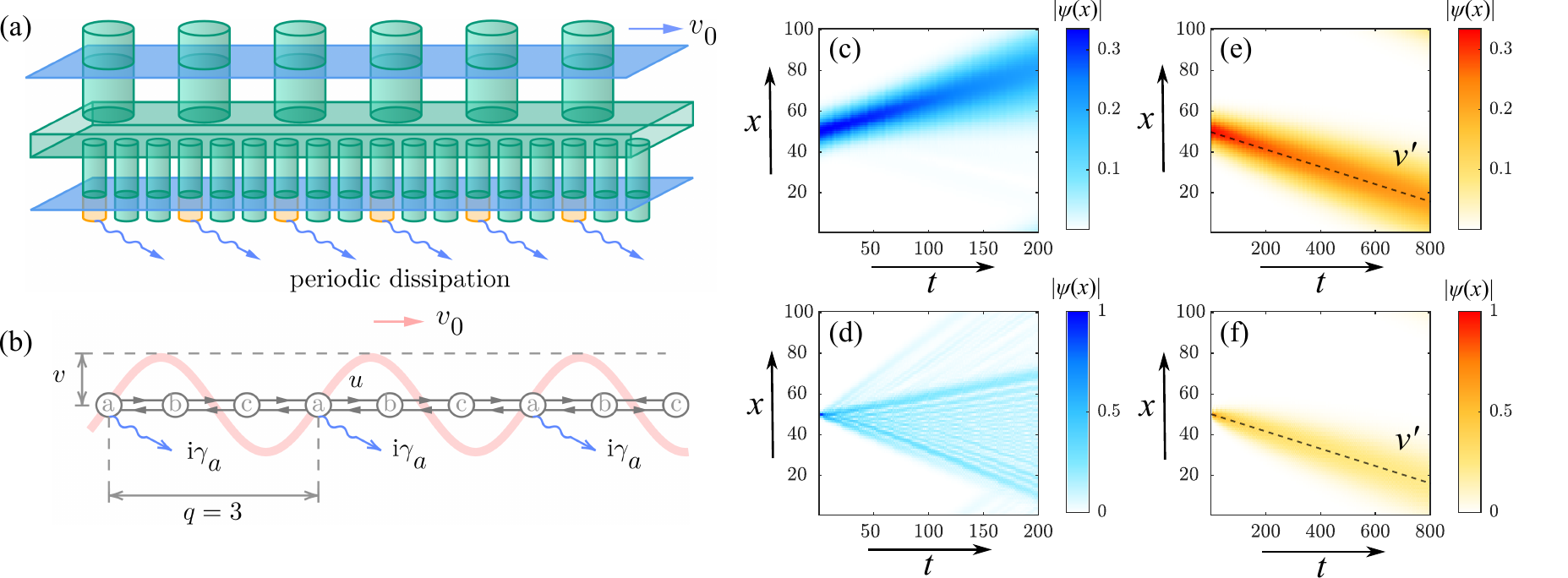}
    \caption{\label{fig:Fig1}
    (a) The illustration of the experimental setup to achieve the periodically driven AAH model with dissipation; 
    (b) The schematic of the tight-binding model. 
    The sinusoidal potential propagates rightward with velocity $v_0$.
    (c)-(f) wavefunction dynamics simulation on the driven AAH periodic chain with parameters $p/q=1/3$, $u=v=1$, system size $N=100$, and driving period $T = 2\pi/\Omega = 5\pi$. 
    The initial wavefunctions are chosen as Gaussian in (c) and (e) and delta function in (d) and (f), centered at $x_0=50$. 
    The corresponding time evolutions without and with loss ($\gamma_a=-1.2$) are shown in (c)(d) and (e)(f), respectively. 
    In (e) and (f), the wavepackets at every time have been renormalized.}
\end{figure*}

To answer this question, we report our discovery of universal one-way transport in the periodically driven Aubry-Andr\'e-Harper (AAH) chain with imbalanced on-site dissipation. 
Our model is closely related to the recent experimental work~\cite{cheng2020experimental} in acoustic waveguides.
In direct contrast to the loss-free case where certain bands exhibit leftward propagation while others rightward, the introduction of damping results in a universal propagation direction for all excitations and wavepackets, which remains robust against any types of impurities. 
We attribute this phenomenon to the emergent non-Hermitian skin effect of the effective Floquet Hamiltonian~\cite{zhang2020FNHSE} and also explored the hidden symmetry that will lead to modes doubling.
Furthermore, we propose to use the time-average decay rate of the propagator as a signature to the non-reciprocal long-time dynamics of the system, which can be probed experimentally. 
Our work sheds light on the implementation of universal one-way transport in a wide range of experimental platforms that AAH-type physics has been achieved~\cite{cheng2020experimental,aah2022atomic,aah2021laser,aah2019photonic}.

\section{Periodically driven AAH model}
The AAH model can be viewed as a 1D reduction of the 2D Hofstadter model~\cite{Bernevig+2013}, which inherits the essential features of 2D quantum Hall systems, including Chern bands and topological edge states~\cite{aah2013}. 
While our focus lies on commensurate AAH models, it is important to note that all key conclusions are equally applicable to incommensurate models~\cite{incommensurate1,incommensurate2,zhou2022driving}  
as shown in the Appendix~\ref{subsec:incom}. 
The AAH model is described by the tight-binding Hamiltonian 
\begin{equation}\label{EQ_StaticAAH}
    \hH_{0}=\sum_{n}\9v\cos(2\pi\phi n+\varphi)\hc_n^\dagger\hc_n+u\,\hc_n^\dagger\hc_{n+1}+\text{h.c.}\0,
\end{equation}
where $\hc_n^\dagger$ ($\hc_n$) labels the creation (annihilation) operator at lattice site $n$, and $u,v$ represent the nearest hopping amplitude and the magnitude of on-site cosine potential, respectively. 
For the on-site potential $v\cos(2\pi\phi n+\varphi)$, $\phi$ determines the spatial period of the potential, and $\varphi$ denotes a phase factor. 
Commensurate models necessitate a rational $\phi=p/q$, where $p$ and $q$ are coprime integers. 
Due to the modulation of the potential, the unit cell is enlarged to $q$ sites per unit cell, thus showing $q$ bands in its spectrum. 

The phase factor in the sinusoidal potential can be designed as time-dependent $\varphi=\varphi(t)=\Omega t$, such that the on-site potential starts to propagate with a velocity $v_0=\Omega/2\pi\phi$. 
Accordingly, the Hamiltonian in Eq.~(\ref{EQ_StaticAAH}) can be rewritten as a time-dependent Hamiltonian
\begin{equation}\label{EQ_DrivenAAH}
    \hH_0(t)=\sum_{n}\9v\cos(2\pi\phi n+\Omega t)\hc_n^\dagger\hc_n+u\,\hc_n^\dagger\hc_{n+1}+\text{h.c.}\0,
\end{equation}
with driving period $T=2\pi/\Omega$. 
The periodically driven AAH model can be realized utilizing bilayered acoustic metamaterials~\cite{cheng2020experimental}, and the related experimental setup is illustrated in Fig.~\ref{fig:Fig1}(a). 
It comprises two layers of acoustic resonators (the green tubes in Fig.~\ref{fig:Fig1}(a)) and an intermediate inner chamber. 
The two layers feature distinct lattice structures, including different tube radii and spacing between neighboring tubes, and can move relative to another layer. 
To realize the driven AAH model represented by Eq.~(\ref{EQ_DrivenAAH}), the top layer needs to move at a constant speed of $v_0$ relative to the bottom layer.

The mode loss is a ubiquitous and inevitable factor and can be flexibly manipulated in the acoustic metamaterial. 
Generally, we can assign imaginary on-site potentials $\rmi\gamma_1$,$\rmi\gamma_2$,$\cdots$,$\rmi\gamma_q$ within the unit cell to simulate the effects of dissipation. 
In this paper, we mainly focus on the case where $\phi = p/q = 1/3$ and the dissipation is only introduced for the first of the three sites of each unit cell, as depicted in Fig.~\ref{fig:Fig1}(a). 
Due to the enlargement of the unit cell, we relabel the basis $\{\hc_0^{\dagger}, \hc_1^{\dagger}, \hc_2^{\dagger},\dots, \hc_{3N-1}^{\dagger} \}$ as $\{\hat{\psi}_{1,a}^{\dagger}, \hat{\psi}_{1,b}^{\dagger}, \hat{\psi}_{1,c}^{\dagger},\dots, \hat{\psi}_{N,c}^{\dagger} \}$. 
Under the new basis, the Hamiltonian contains $N$ unit cells and $3$ sites per unit cell and the total time-dependent Hamiltonian in momentum space is given by $\hH = \hat{\psi}_k^{\dagger}\mathcal{H}(k,t)\hat{\psi}_k$, where $\hat{\psi}_k = (\hat{\psi}_{k,a},\hat{\psi}_{k,b},\hat{\psi}_{k,c})^T$ and 
\begin{equation}\label{EQ_LossDrivenAAH}
\begin{split}
\cH=\mqty(v\cos\Omega t+\rmi\gamma_a&-u&-u\rme^{-\rmi k}\\-u&v\cos(\Omega t+\frac{2\pi}{3})&-u\\-u\rme^{\rmi k}&-u&v\cos(\Omega t+\frac{4\pi}{3}))
\end{split}
\end{equation}
with $\gamma_a$ the only non-Hermitian parameter. 

\begin{figure*}[t]
    \centering
    \includegraphics[width=1\linewidth]{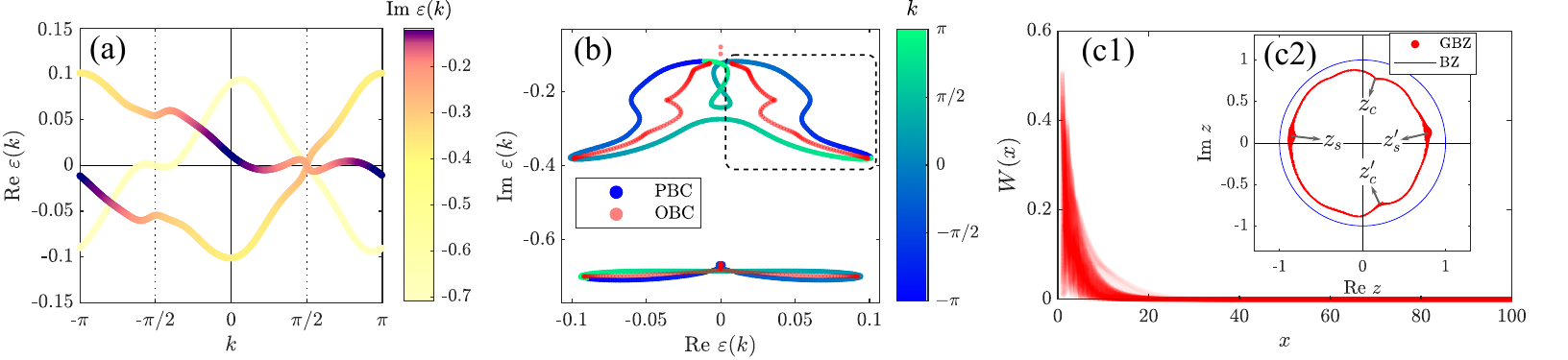}
    \caption{\label{fig:Fig2}
    (a) The real part of quasienergy band structures is weighted by the corresponding imaginary energy (indicated by the color bar); 
    (b) The green-blue curve indicates the quasienergies under periodic boundary conditions (PBC), where $k$ is labeled by the color bar. 
    The red dots represent the eigenvalues under open boundary conditions (OBC); 
    (c1) Numerically calculated GBZ for the boxed non-Bloch quasienergy band in (b). 
    The GBZ of this band is totally inside BZ, indicating the non-Hermitian skin modes are totally localized at the left edge as shown in (c2).
    The parameters are the same as in Fig.~\ref{fig:Fig1}.}
\end{figure*}

\section{Loss-induced universal one-way transport}
Now, we will demonstrate the one-way transport via wavepacket dynamics at the level of stroboscopic evolution. 
Hence, the time scale considered in the following simulation is much larger than the driving period $T=2\pi/\Omega$. 
The stroboscopic dynamics is governed by the effective Floquet Hamiltonian $H_F(k)$ defined as 
\begin{equation}\label{EQ_FloquetTrans}
    U(k,T)=\hat{\cT}\rme^{-\rmi\int_0^T \cH(k,t)\dd t}\equiv\rme^{-\rmi H_F(k) T},
\end{equation}
where $\hat{\cT}$ represents the time ordering operator. 

For comparison, we first present the wave propagation in a loss-free AAH chain ($\gamma_a=0$). 
At time $t=0$, we prepare a Gaussian wavepacket centered at $x_0$ as the initial condition: $\bm{\psi}(x,t=0)=\cN\exp[-(x-x_0)^2/2\sigma^2]\,\bm{u}$ where $\cN$ indicates the normalization factor and $\sigma$ represents the width of the wavepacket. $\bm{u}$ is a three-component vector and here we set it to match the Bloch wavefunction of the $\alpha$-th energy band at $k=0$.
The wavefunction at any given time is obtained by numerically evaluating $\bm{\psi}(x,t)=\zf\hat{\cT}\exp(-\rmi\int_{0}^t H(\tau)\dd \tau)\yf\bm{\psi}(x,0)$, where $H(\tau)$ denotes the instantaneous real-space Hamiltonian at time $\tau$.
Without driving ($\Omega=0$), this wavepacket will disperse symmetrically in both leftward and rightward directions, due to the time-reversal symmetry. Under periodic driving ($\Omega\neq 0$), time-reversal symmetry is broken, and the quasienergy dispersion of $H_F(k)$ is tilted with nonzero group velocity $\partial_k \varepsilon_{\alpha}(k=0)\ne 0$, and thus the wavepacket will start to propagate [Fig.~\ref{fig:Fig1}(c)]. However, this propagation direction is not universal, but depends on the choice of bands ($\alpha$). In fact, for this effective Floquet Hamiltonian, one can prove that
at any $k$, the sum of velocities for all quasienergy bands must be zero: $\sum_{\alpha=1}^q v_j=\partial_k \Tr H_F(k)=0$. Therefore, any propagating mode must always be accompanied by counter-propagating modes, fully consistent with the no-go theorem~\cite{sun2018chiralmagnetic}.
Because modes can propagate in both directions, wavepackets in the loss-free model can be backscattered to other propagation channels by impurities (see Appendix~\ref{ap:robust}), indicating the lack of universal one-way transport.
To further demonstrate these counter-propagating modes, we replace the Gaussian wavepacket with a delta function $\bm{\psi}(x,t=0)=\delta(x-x_0)$ initial condition, which will excite all eigenmodes of $H_F$. As shown in Fig.~\ref{fig:Fig1}(d), one-way transport becomes totally unattainable here due to the presence of counter-propagating modes.

Remarkably, by simply introducing some dissipation $\gamma_a<0$, all wavepackets will now uniformly travel in the $-x$ direction at the same velocity $v'$, regardless of their initial wavefunctions, as depicted in Fig.~\ref{fig:Fig1}(e)(f).
Furthermore, this unidirectional propagation is robust against impurity scatterings, as shown in Appendix~\ref{ap:robust}. 
This implies that dissipation triggers universal one-way transport, which is one of the key observations in this study.
Given the experimental feasibility of manipulating dissipation, we consequently pose the following crucial questions:
What are the requirements for the emergence of backscattering-free one-way transport induced by dissipation? 
Is there an underlying mechanism behind the observed phenomenon? 

\section{The role of symmetry}
Here, we reveal that the observed robust one-way transport is guaranteed by the symmetry hidden in the effective Floquet Hamiltonian.

We start from the $q=2$ AAH model where the time-dependent Hamiltonian in Eq.~(\ref{EQ_LossDrivenAAH}) reduces to a $2*2$ matrix and respects the symmetry $\cH^T(-k,-t) = \cH(k,t)$. 
Translating the symmetry to  $H_F$ using Eq.~(\ref{EQ_FloquetTrans}), we get 
\begin{equation}\label{EQ_Q2Symmetry}
    H_{F,q=2}^T(-k) = H_{F,q=2}(k),
\end{equation}
which implies the effective Floquet Hamiltonian with $q=2$ is reciprocal and prohibits non-reciprocal transport~\cite{Kawabata2020PRB}. 

The symmetry constraint in Eq.~(\ref{EQ_Q2Symmetry}) is relieved for the $q\geq 3$ models. 
Generally, the instantaneous Hamiltonian $\cH(k,t)$ in Eq.~(\ref{EQ_LossDrivenAAH}) contains $q$ bands and respects the following symmetry:
\begin{equation}
    \cU^{\dagger} \cH^{*}(k,t) \cU = -\cH\9 q\pi-k,t+\frac{T}{2}\0
    \label{eq:model_symmetry},
\end{equation}
where the unitary transformation matrix $\cU$ is found to be $\cU_{ij}=(-1)^{i+1} \delta_{ij}$. 
We can further obtain the symmetry of the effective Floquet Hamiltonian $H_F$:
\begin{equation}\label{EQ_EffectiveSymm}
\mathcal{V}^\dagger H_F^{\ast}\9k\0\mathcal{V}=-H_F\9q\pi-k\0
\end{equation}
with $\mathcal{V}=\cU\9\hat{\cT}\rme^{-\rmi\int_0^{T/2}\cH^*(\pm q\pi-k,\tau)\dd\tau}\0$. 
Therefore, the complex quasienergies as eigenvalues of $H_F$ satisfy 
\begin{equation}\label{EQ_QuasiEngSymm}
    \varepsilon(k) = - \varepsilon^{\ast}(q\pi - k).
\end{equation}
It shows that the real (imaginary) part of the quasienergy bands is antisymmetric (symmetric) about $k=q \pi/2$ and $q \pi/2-\pi$. 
For example, the quasienergy bands with $q=3$ are depicted in Fig.~\ref{fig:Fig2}(a), where the real part $\Re\,\varepsilon(k)$ is an odd function about $k=\pm\pi/2$, whereas $\Im\,\varepsilon(k)$ is even about the same points. 

Upon long-time evolution, the dynamics is governed by the quasienergies with the largest imaginary part (the lowest damping).
Suppose that the maximal imaginary quasienergy $\Im\,\varepsilon(k_m)$ is attained at $k_m$.
According to the relation in Eq.~(\ref{EQ_QuasiEngSymm}), there must be $k^{\prime}_m = \pi-k_m$ possessing the same imaginary part and the identical velocity $\partial_{k}\Re\,\varepsilon(k_m) = \partial_{k}\Re\,\varepsilon(k^{\prime}_m)$. 
Hence, the symmetry Eq.~(\ref{EQ_EffectiveSymm}) of the effective Floquet Hamiltonian guarantees unidirectional transport under long-time evolution, regardless of the details in the initial excitation. 
As an example, the velocity $v'$ of the wavepackets in Fig.~\ref{fig:Fig1}(e)(f) coincides with the slope of the segment of the quasienergy band with the largest imaginary part in Fig.~\ref{fig:Fig2}(a), namely $v'=\partial_k\Re\,\varepsilon(k_m)$.

\section{Floquet emergent non-Hermitian skin effect and stroboscopic generalized Brillouin zone}
Here, we elucidate the underlying fundamental mechanism driving the universal one-way transport, associated with the emergent non-Hermitian skin effects in the effective Floquet Hamiltonian. 

We examine the eigenstates and eigenvalues of the effective Floquet Hamiltonian $H_F$, as these two dictate the system's stroboscopic dynamics.
The eigenvalues of the effective Floquet Hamiltonian $H_{F}$ are numerically obtained due to the lack of its analytical expression.
The periodic-boundary spectra of $H_F$ form some loops with nonzero interiors, as plotted by the blue dots in Fig.\ref{fig:Fig2} (b), which is totally different from the open-boundary eigenvalues indicated by the red dots in Fig.\ref{fig:Fig2} (b). 
This spectral difference signals the existence of non-Hermitian skin effects~\cite{Kai2020,Okuma2020}. 
As a verification, we calculate the spatial distribution of the open-boundary wavefunctions
\begin{equation}\label{EQ_SpatialWave}
    W(x) = \sum\nolimits_{\alpha=1}^q|\varphi_{\alpha}(x)|^2,
\end{equation}
where $\varphi_{\alpha}(x)$ represents the eigenfunction of $H_F$ and $q=3$ in our model, plotted by the red lines in Fig.~\ref{fig:Fig2}(c1). 
It shows that the open-boundary wavefunctions are localized at the left edge, in accordance with the leftward propagation of wavepackets in Fig.~\ref{fig:Fig1}(e)(f). 
Note that for any given time $t$, the instantaneous Hamiltonian in Eq.~(\ref{EQ_LossDrivenAAH}) respects reciprocity, $\cH^T(-k,t) = \cH(k,t)$, thus lacking the non-Hermitian skin effect~\cite{YiYang2020PRL,Kawabata2020PRB}. 
Actually, the non-Hermitian skin effect emerges from the periodic driving, termed Floquet emergent non-Hermitian skin effect. 

\begin{figure}[b]
    \centering
    \includegraphics[width=1\linewidth]{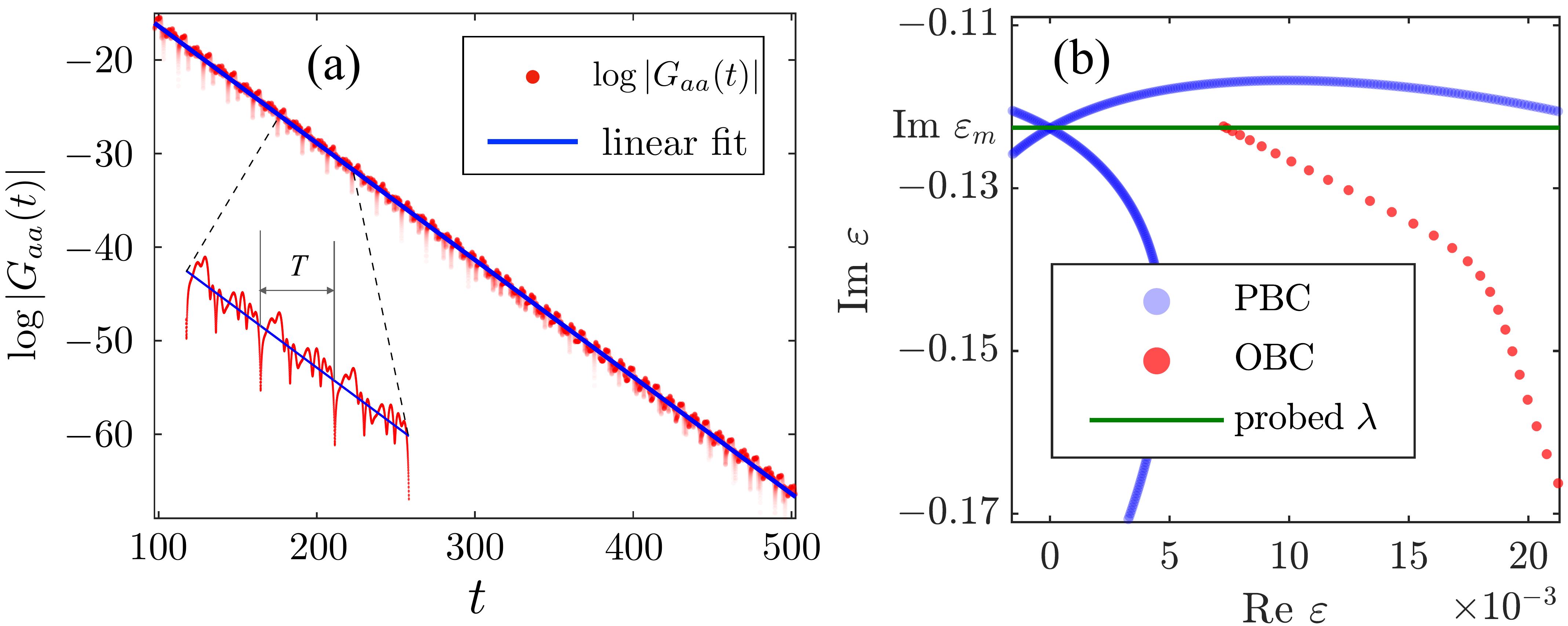}
    \caption{Long-time dynamical feature. 
    (a) The linear fitting for the decay of $\log|G_{aa}(t)|$. A segment is zoomed in to show the periodic oscillation; 
    (b) Comparing the fitted slope with the largest imaginary part of the OBC spectrum. The OBC and PBC eigenvalues are a local magnification of Fig.~\ref{fig:Fig2}(b). 
    Model parameters are the same as Fig.~\ref{fig:Fig1}.}
    \label{fig:Fig3}
\end{figure}

To further investigate the features of $H_F$, we numerically calculate the corresponding generalized Brillouin zone (GBZ)~\cite{Yao2018,Murakami2019,yang2020aGBZ}. 
In a band system with Bloch Hamiltonian $\cH(k)$, the GBZ can be obtained by extending the momenta $k$ into complex $k-\rmi \mu(k)$ such that $\cH(k-\rmi\mu(k))$ reproduces the open-boundary spectrum, where $1/\mu(k)$ represents the decay length and generally varies with $k$. 
However, the effective Hamiltonian $H_F$ lacks an analytic expression for which we adopt the following method.
The characteristic polynomial of $H_F$ can be expressed as $f(k,E) = \det [H_F(k) - E \mathbb{I}_3]$. For a given open-boundary eigenvalue $E_0$, we search for the $k$ and $\mu$ such that $f(k-\rmi\mu,E_0)=0$~(More details are presented in Appendix~\ref{ap:FGBZ}).
Finally, we obtain $\mu(k,E_0)$ for $E_0$ the open-boundary eigenvalue. 
We plot the GBZ as a curve of $z:=\rme^{\rmi(k-\rmi\mu(k,E))}$ for one of the open-boundary quasienergy bands in Fig.~\ref{fig:Fig2}(c2). 
It shows that the GBZ is totally included inside BZ (the blue unit circle), therefore, the eigenstates on the quasienergy band shown in the black box in Fig.~\ref{fig:Fig2}(b) are localized at the left edge, consistent with the numerical results in Fig.~\ref{fig:Fig2}(c1). 
Our calculation captures two types of feature points on the GBZ labeled as $z_c$ and $z_s$ in Fig.~\ref{fig:Fig2}(c2). 
The two points $z_c,z_c^{\prime}$ come from the cusp on the quasienergy band. 
Especially, the two saddle points $z_s$ and $z_s^{\prime}$ correspond to the two endpoints of the quasienergy band in the black dashed box in Fig.~\ref{fig:Fig2}(b). 
Due to the high spectral density at two endpoints, the numerically calculated GBZ appears denser at these two saddle points, as shown in Fig.~\ref{fig:Fig2}(c2). 

\section{Long-time dynamical feature}

Based on the information from GBZ, we predict the time-average decay rate of the propagator as a long-time dynamical feature of Floquet emergent non-Hermitian skin effect, which can be directly probed in experiments~\cite{XuePeng2021PRL}. 
The on-site element of the propagator deep in the bulk at a random site $a$ is
\begin{equation}\label{EQ_LyapunovExp}
 G_{aa}(t)=\mel{a}{\hat{\mathcal{T}}\mathrm{e}^{-\mathrm{i}\int_0^t \hat{H}(\tau)\dd \tau}}{a}.
\end{equation}
After a sufficiently long time, it will reach a steady state with exponential decay, modulated by periodic oscillations.
Using stationary phase approximation, we can prove the time-average decay rate, defined as
\begin{equation}
    \lambda = \overline{\partial_t\log|G_{aa}(t)|}
\end{equation}
where the overline denotes time-averaging,
is determined by the saddle points energy with the largest imaginary part (see more details in Appendix~\ref{ap:LongTime}). 
In our model, it happens to be the endpoint of the open-boundary quasienergy band with the largest imaginary part $\Im\,\varepsilon_m$, which corresponds to the saddle points $z_s,z_s^{\prime}$ on the GBZ in Fig.~\ref{fig:Fig2}(c2).
This quantity, termed the Lyapunov exponent, has been discussed in static band systems~\cite{longhi2019probing}.

In our simulation, we plot the $\log|G_{aa}(t)|$ by the red dots in Fig.~\ref{fig:Fig3}(a).
The blue line is the fitted line with slope $\lambda$, which is approximately in accordance with the maximal imaginary part of the open-boundary quasienergies $\Im \, \varepsilon_m$, as shown in Fig.~\ref{fig:Fig3}(b). 
Note that the $\log|G_{aa}(t)|$ exhibits periodic oscillation and overall decay rate $\lambda$, which is distinct from the situation in band systems~\cite{longhi2019probing}. 
In experiments, one can just probe the on-site propagator at site $a$ and record the decay rate to compare with $\Im \varepsilon_m$, which is a clear signature of the Floquet emergent non-Hermitian skin effect. 

\section{Conclusion}
In summary, we propose that the imbalanced on-site gain/loss enables universal one-way transport in a periodically driven AAH chain, which is independent of initial excitations and immune to impurity scatterings. 
Given that dissipation is commonly present in realistic systems, we expect that the loss-induced universal one-way transport can be achieved in a wide range of experimental platforms. 
We show the appearance of this phenomenon is ensured by symmetries of the effective Floquet Hamiltonian and can be related to the Floquet emergent non-Hermitian skin effect that is absent in the instantaneous Hamiltonian. 
We use the time-averaged decay rate of the propagator
, an experimentally measurable quantity, as the long-time dynamical feature to probe the Floquet emergent non-Hermitian skin effect. 

\section*{Acknowledgements}
We thank N. Cheng and W. Cheng for valuable discussions. 
This work was supported in part by the Office of Naval Research (MURI N00014-20-1-2479). 

\appendix
\section{Introduction of the AAH model and its Floquet variant}
\begin{figure*}[t]
    \centering
    \includegraphics[width=0.7\linewidth]{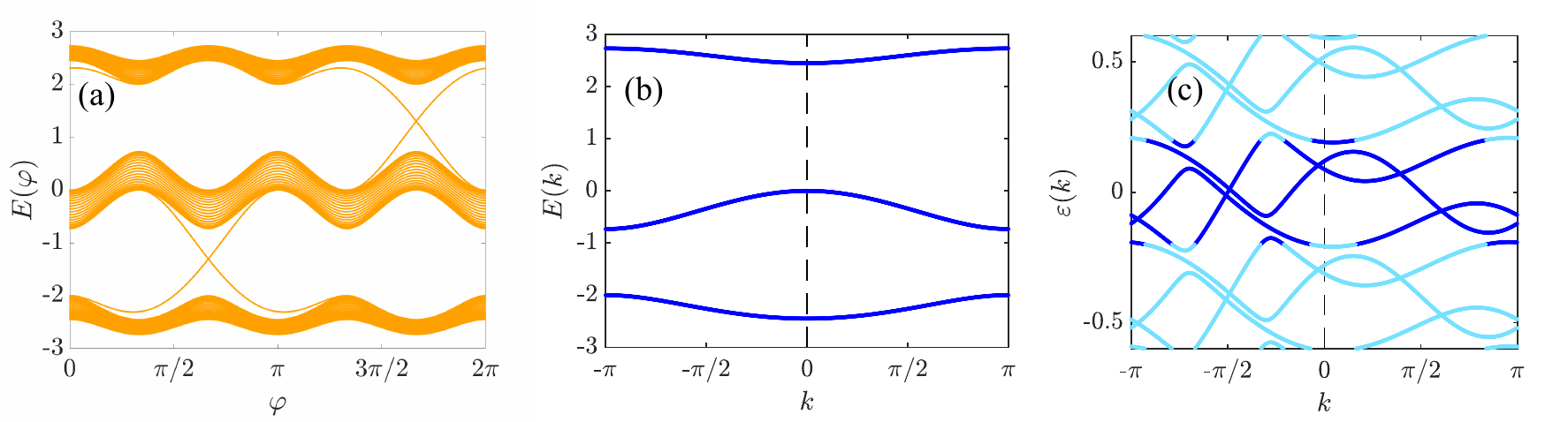}
    \caption{(a) The energy spectrum of a family of $q=3$ AAH chain under open boundary condition as a function of the effective magnetic flux $\varphi$. Chiral edge states can be observed;
    (b) The energy spectrum of a $q=3$ AAH chain under periodic boundary condition as a function of reciprocal vector $k$;
    (c) When Floquet driving is turned on, the quasienergies spectrum of the Floquet $q=3$ AAH chain.
    The darker blue indicates the minimum periodic repetitive unit of quasienergy spectrum.}
    \label{fig:SF0}
\end{figure*}

In this section, we offer a brief introduction to the AAH model and the Floquet version of it. 
The detailed introduction can be found in, for example, Chap.~5 of Ref.\cite{Bernevig+2013}. 
An important insight is to view this model as a reduction of the 2D Hofstadter model to 1D. 
The 2D Hofstadter model describes a 2D square lattice in a uniform perpendicular magnetic field. 
Using Landau gauge and Peierls substitution, the 2D Hamiltonian is given by
\begin{equation}
 \hH_{\text{2D}}=-t\sum_{ij}\9\hc^{\dagger}_{i+1,j}\hc_{i,j}+\hc^{\dagger}_{i,j+1}\hc_{i,j}\rme^{2\pi\rmi\phi}+\text{h.c.}\0,
\end{equation}
where $\{i,j\}$ label the lattice sites, and $\phi$ denotes the flux per unit cell (in the unit of flux quantum $h/e$). 
Now, we restrict our discussion to commensurate models, where the flux number $\phi=p/q,\ p,q\in\mathbb{Z}$. 
The dynamics of the incommensurate model will be addressed in section \ref{subsec:incom}. 
Because of the fractional magnetic flux, the unit cell is enlarged by $q$ times in the $x$ direction. 
After the Fourier transform, the Hamiltonian in reciprocal space is given by 
\begin{equation}
\begin{aligned}
    \hH(k_x,k_y)=-t\sum_{n=0}^{q-1}(\cos(2\pi\phi n+k_x)\hc^\dagger_{n}\hc_{n}\\+\rme^{\rmi k_y}\hc^\dagger_{n}\hc_{n+1}
    +\text{h.c.}).
    \label{eq:2Dham_k}
\end{aligned}
\end{equation}
where $\hc_n$ as a short notation for $\hc_{k_x+2\pi\phi n,k_y}$. Now, map the Hamiltonian (\ref{eq:2Dham_k}) to the 1D AAH model and convert $n$ to a genuine 1D lattice label.
\begin{equation}
     \hH_{1D}=\sum_{n}\9v\cos(2\pi\phi n+\varphi)\hc_n^\dagger\hc_n+u\hc_n^\dagger\hc_{n+1}+\text{h.c}\0.
\end{equation}
where we have absorbed $u=-t\rme^{\rmi k_y}$ and $\varphi=k_x$. This model is just a tight-binding chain with hopping amplitude $u$ modulated by a cosine on-site potential with amplitude $v$ (from the exact reduction $u$ and $v$ must have the same magnitude, but altering $v$ does not affect the magnetic translation algebra. So we lift this constraint of $u$ and $v$ for general purpose). For $\phi=p/q$, this model contains $q$ sites within a unit cell, hence having $q$ bands in its spectrum. Since it is the descendent of the 2D Hofstadter model, it inherits  some of its essential properties including bands with nonzero Chern number and topological edge states. For $q$ even, it can be proven that the spectrum contains $q$ Dirac cones between the middle two bands \cite{WEN1989641}. If $q=2$, this model is topologically trivial as it can be mapped to the 2D $\pi$-flux model \cite{aah2013}. If $q\geq4$, the Chern number for the middle two bands is $q-2$, canceling the Chern number of other bands, each carrying Chern number $-1$. For $q$ odd and $q\geq 3$, only one middle band carries a Chern number of $q-1$, canceling the Chern number of other bands with Chern number $-1$. Therefore, under open boundary conditions, topological edge states can be observed within the gap as shown in Fig.~\ref{fig:SF0}(a). 

In the main text, we consider the Floquet version of the AAH model, where the phase factor $\varphi$ becomes time-dependent $\varphi(t)=\Omega t$. When $t$ scans from $0$ to $T=2\pi/\Omega$, it is equivalent to scanning $k_x$ from $0$ to $2\pi$ in the 2D Hofstadter model. Notice that the static AAH model has time-reversal symmetry  $E(k)=E(-k)$. Hence, the slopes at $k=0,\pm\pi$ are all zero. If one creates excitations around $k=0,\pm\pi$, the excitations will disperse without drifting (Fig.~\ref{fig:SF0}(b)). Adding periodic driving breaks the time-reversal symmetry and therefore the quasienergy bands are tilted at $k=0,\pm\pi$, and the excitations will propagate according to the slopes of the quasienergy bands (Fig.~\ref{fig:SF0}(c)).

Although periodic driving makes excitations around a certain $k$ propagate according to the slopes or velocities of the corresponding Floquet bands, they cannot lead to unidirectional transport. Notice that in this model, the diagonal elements of $\cH$ is $k$-independent (except for the $q=1$ case, but in this case the Floquet band is trivial). Hence, the addition of the velocities of the $q$ bands gives
\begin{equation}
    \sum_{i=1}^q v_i\equiv\sum_{i=1}^q\frac{\partial \varepsilon_i(k)}{\partial k}=\frac{\partial}{\partial k}\Tr{H_F(k)}=0.
\end{equation}

\section{The role of symmetry}
In this section, we will present the details of the proof of the model's symmetries in the main text.   Also, we will include detailed discussions about the mode doubling in the system due to the hidden symmetry.

\textbf{Reciprocity of the \bm{$q=2$} model:} When $q=2$, the time-dependent Hamiltonian is given by
\begin{equation}
    \cH(k,t)=\mqty(v\cos\Omega t+\rmi\gamma&-u-u\rme^{-\rmi k}\\-u-u\rme^{\rmi k}&-v\cos\Omega t).
\end{equation}
It is easy to verify that it satisfies the following symmetry:
\begin{equation}
    \cH(k,t)=\cH^T (-k,-t).
    \label{eq:q=2_sym}
\end{equation}
Next, we translate this symmetry to the symmetry of the effective Floquet Hamiltonian $ H_F$. Set $\delta t=T/N$ and $t_i=i\cdot\delta t,\ i=1,2,\cdots,N$, then we have
\begin{equation}
    \rme^{-\rmi H_F(k) T}\equiv\hat{\cT}\rme^{-\rmi\int_{0}^{T}\cH(k,\tau)\dd\tau}=\lim_{N\to\infty}\rme^{-\rmi\cH(k,t_N)\delta t}\cdots\rme^{-\rmi\cH(k,t_1)\delta t}.
\end{equation}
Transpose both sides and we get
\begin{equation}
    \rme^{-\rmi H_F^T(k) T}=\lim_{N\to\infty}\rme^{-\rmi\cH^T(k,t_1)\delta t}\cdots\rme^{-\rmi\cH^T(k,t_N)\delta t}.
\end{equation}
Using the symmetry (\ref{eq:q=2_sym}), we get
\begin{equation}
    \rme^{-\rmi H_F^T(k) T}=\rme^{-\rmi H_F(-k)T}.
\end{equation}
Therefore, the symmetry of the effective Hamiltonian is
\begin{equation}
     H_F(k)= H_F^T(-k)
\end{equation}
which enforces that the $q=2$ model is reciprocal.~\cite{Okuma2020}

\textbf{The general symmetry for all \bm{$q$}:} In the main text, the symmetry of the Floquet AAH model with dissipation is given by Eq.~(\ref{eq:model_symmetry}). Here we rewrite this important symmetry
\begin{equation}
    \cU^{\dagger} \cH^{*}(k,t) \cU = -\cH\9\pm q\pi-k,t+\frac{T}{2}\0.
    \label{eq:model_sym}
\end{equation}
For the $q=1$ case, this symmetry can be directly verified. For the $q\geq 2$ case, the time-dependent Hamiltonian $\cH(k,t)$ under periodic boundary conditions is given by
\begin{widetext}
\begin{equation*}
\cH(k,t)=\mqty(v\cos(\Omega t)+\rmi\gamma&-u&0&\cdots&-u\rme^{-\rmi k}\\
-u&v\cos(\Omega t+\frac{1}{q}\cdot 2\pi)&-u&\cdots&0\\
0&-u&v\cos(\Omega t+\frac{2}{q}\cdot 2\pi)&\cdots&0\\
\vdots&\vdots&\vdots&\ddots&\vdots\\
-u\rme^{\rmi k}&0&0&\cdots&v\cos(\Omega t+\frac{q-1}{q}\cdot 2\pi)).
\end{equation*}
\end{widetext}
We claim that the unitary transformation matrix $\cU$ in the symmetry (\ref{eq:model_sym}) is
\begin{equation}
    \cU=\operatorname{diag}\{1,-1,1,\cdots,(-1)^{q+1}\}
\end{equation}
After performing this unitary transformation, the Hamiltonian becomes
\begin{widetext}
\begin{equation}
    \cU^\dagger\cH^*(k,t) \cU=\mqty(v\cos(\Omega t)-\rmi\gamma&u&0&\cdots&(-1)^q u\rme^{\rmi k}\\
u&v\cos(\Omega t+\frac{1}{q}\cdot 2\pi)&u&\cdots&0\\
0&u&v\cos(\Omega t+\frac{2}{q}\cdot 2\pi)&\cdots&0\\
\vdots&\vdots&\vdots&\ddots&\vdots\\
(-1)^q u\rme^{-\rmi k}&0&0&\cdots&v\cos(\Omega t+\frac{q-1}{q}\cdot 2\pi)).
\label{eq:sym_LHS}
\end{equation}
Meanwhile, the right-hand side of (\ref{eq:model_sym}) is given by
\begin{equation}
    \cH\9\pm q\pi-k,t+\frac{T}{2}\0=\mqty(-v\cos(\Omega t)+\rmi\gamma&-u&0&\cdots&(-1)^{q+1}u\rme^{\rmi k}\\
-u&-v\cos(\Omega t+\frac{1}{q}\cdot 2\pi)&-u&\cdots&0\\
0&-u&-v\cos(\Omega t+\frac{2}{q}\cdot 2\pi)&\cdots&0\\
\vdots&\vdots&\vdots&\ddots&\vdots\\
(-1)^{q+1}u\rme^{-\rmi k}&0&0&\cdots&-v\cos(\Omega t+\frac{q-1}{q}\cdot 2\pi)).
\label{eq:sym_RHS}
\end{equation}
\end{widetext}
By comparing~(\ref{eq:sym_LHS}) and~(\ref{eq:sym_RHS}), we verified the symmetry~(\ref{eq:model_sym})
\begin{equation}
    \cU^{\dagger} \cH^{*}(k,t) \cU = -\cH\9\pm q\pi-k,t+\frac{T}{2}\0.
\end{equation}

\textbf{Insensitivity to the dissipation distribution}: In the main text, we considered the model $q=3$ and only added dissipation for the first of the three sites of a unit cell. Actually, we can show that we can actually randomly distribute the gain/loss within the unit cell without changing the model symmetries. If a random distribution $\cH_{\text{dis}}=\text{diag}\{\rmi\gamma_1,\rmi\gamma_2,\cdots,\rmi\gamma_q\}$ is added, then
\begin{equation}
    \cU^\dagger \cH^*_{\text{dis}} \cU= -\cH_{\text{dis}}
\end{equation}
regardless of whether $q$ is even or odd. Therefore, the symmetry~(\ref{eq:model_sym}) still holds for $\cH(k,t)=\cH_0(k,t)+\cH_{\text{dis}}$.
Practically, this means that the phenomena can be observed using gain as well. Moreover, we could add uniform gain/loss on top of the original model to globally shift the whole spectrum vertically in the complex plane so that the highest segments have zero imaginary parts. Thus, the one-way propagation will persist without decay or amplification.

\textbf{Symmetry of the effective Hamiltonian}: Now we extend the symmetry (\ref{eq:model_sym}) to the symmetry of $ H_F$ given by Eq.~(\ref{EQ_EffectiveSymm}) in the main text. Set $\delta t=T/N$ and $t_i=i\cdot\delta t,\ i=1,2,\cdots,N$, then the effective Hamiltonian can be expressed as
\begin{equation}
    \rme^{-\rmi H_F(k) T}=\lim_{N\to\infty}\rme^{-\rmi\cH(k,t_N)\delta t}\cdots\rme^{-\rmi\cH(k,t_1)\delta t}.
\end{equation}
After taking the complex conjugate and performing the unitary transformation, we get
\begin{equation}
    \cU^\dagger\rme^{\rmi H_F^*(k)T}\cU=\lim_{N\to\infty}\cU^\dagger\rme^{\rmi\cH^*(k,t_N)\delta t}\cdots\rme^{\rmi\cH^*(k,t_1)\delta t}\cU
\end{equation}

\begin{figure*}[t]
    \centering
    \includegraphics[width=0.6\linewidth]{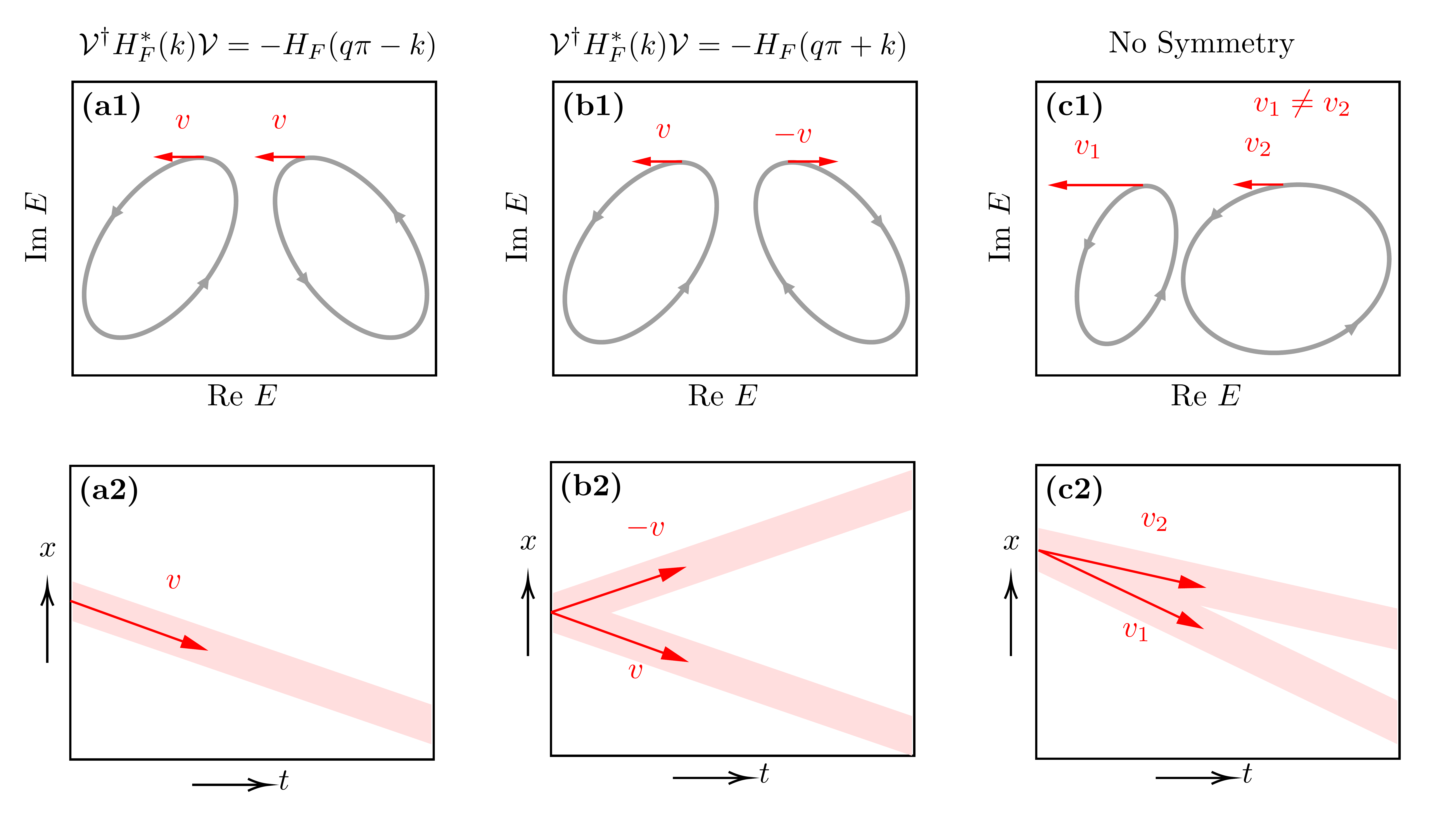}
    \caption{The illustration of the role of the hidden symmetry (Eq.~(6) and Eq.~(7) in the main text). (a1)(a2) shows that with the hidden symmetry Eq.~(7) present, the mode with the largest imaginary part is doubled with the same velocity component, which is exactly what happens in the Floquet AAH model. (b1)(b2) show that if the symmetry is modified to \ref{eq:minus_sym}, we will have a velocity component with opposite propagation direction. (c1)(c2) shows that without any symmetry constraint, the modes with the largest imaginary part can have different group velocities.}
    \label{fig:roleofsym1}
\end{figure*}

The unitary transformation can be taken to the exponential, and then we get
\begin{equation}
    \rme^{\rmi \cU^\dagger H^*_F(k)\cU T}=\lim_{N\to\infty}\rme^{\rmi \cU^\dagger\cH^*(k,t_N)\cU\delta t}\cdots\rme^{\rmi \cU^\dagger\cH^*(k,t_1)\cU\delta t}.
\end{equation}
Using (\ref{eq:model_sym}), we get
\begin{equation}
\begin{aligned}
    &\rme^{\rmi \cU^\dagger H^*_F(k)\cU T}\\=&\lim_{N\to\infty}\rme^{-\rmi \cH\9\pm q\pi-k,t_N +\frac{T}{2}\0\delta t}\cdots\rme^{-\rmi \cH\9\pm q\pi-k,t_1 +\frac{T}{2}\0\delta t}.
\end{aligned}
\end{equation}
The left-hand side of the equation is just the evolution operator of one period, but the origin of time is shifted by $T/2$. Another unitary transformation $\mathcal{S}$ can be applied to fix the shifted origin:
\begin{equation}
\begin{aligned}
    &\mathcal{S}^\dagger\rme^{\rmi \cU^\dagger H^*_F(k)\cU T}\mathcal{S}\\=&\lim_{N\to\infty}\rme^{-\rmi \cH\9\pm q\pi-k,t_N\0\delta t}\cdots\rme^{-\rmi \cH\9\pm q\pi-k,t_1\0\delta t}
\end{aligned}
\end{equation}
where $\mathcal{S}$ is given by
\begin{equation}
\mathcal{S}=\hat{\cT}\rme^{-\rmi\int_{0}^{T/2}\cH(\pm q\pi-k,\tau)\dd\tau}
\end{equation}
It follows that
\begin{equation}
\begin{aligned}
    \rme^{\rmi(\cU \mathcal{S})^\dagger H_F(k)(\cU \mathcal{S})T}=&\hat{\cT}\rme^{-\rmi\int_{0}^T\cH(\pm q\pi-k,\tau)\dd\tau}\\=&\rme^{-\rmi H_F(\pm q\pi-k)T}.
\end{aligned}
\end{equation}     
Comparing the expression on the exponential and we finally reach the expression for the symmetry of the effective Floquet Hamiltonian $ H_F$
\begin{equation}
    \cV^\dagger H^*_F(k)\cV=- H_F(\pm q\pi-k)
\end{equation}
where the unitary transformation for $ H_F$ is given by
\begin{equation}
    \cV=\cU\mathcal{S}=\cU\9\hat{\cT}\rme^{-\rmi\int_{0}^{T/2}\cH(\pm q\pi-k,\tau)\dd\tau}\0.
\end{equation}

\textbf{The doubling of velocity modes: }Actually, this hidden symmetry leads to the doubling of modes with the largest imaginary part.
In the main text, we proved the modes with the largest imaginary part must be paired and have the same group velocity. 
Generally speaking, if we have two modes with the largest imaginary parts, they do not necessarily own the same group velocity.
Hence, the long-time dynamics of the system admits more than one velocity component as shown in Fig.~\ref{fig:roleofsym1}.
The hidden symmetry of the periodically driven AAH model ensures the propagation is strictly one-way as there is only one velocity component left. (Fig.~\ref{fig:roleofsym1} (a2))
If this symmetry is modified to 
\begin{equation}
    \mathcal{V}^{\dagger} H_F^*(k) \mathcal{V}=-H_F(q \pi+k)
    \label{eq:minus_sym}
\end{equation}
Then we can still have Floquet non-Hermitian skin effect due to non-zero spectral winding, but the modes with the largest imaginary part will be paired with opposite group velocity. (Fig.~\ref{fig:roleofsym1} (b1)(b2))
This means the propagation will be bidirectional and can be back-scattered by impurities.

The doubling of the velocity modes can have potential applications in transmitting information.
If we label one mode as ``0'' and another as ``1'', then we could encode information into the doubled modes that will be sent to the receiver at the same time with the same amplitudes that are exponentially larger than noise.
This could be a robust method of sending information unidirectionally in a lossy environment.

\begin{figure}[h]
    \centering
    \includegraphics[width=\linewidth]{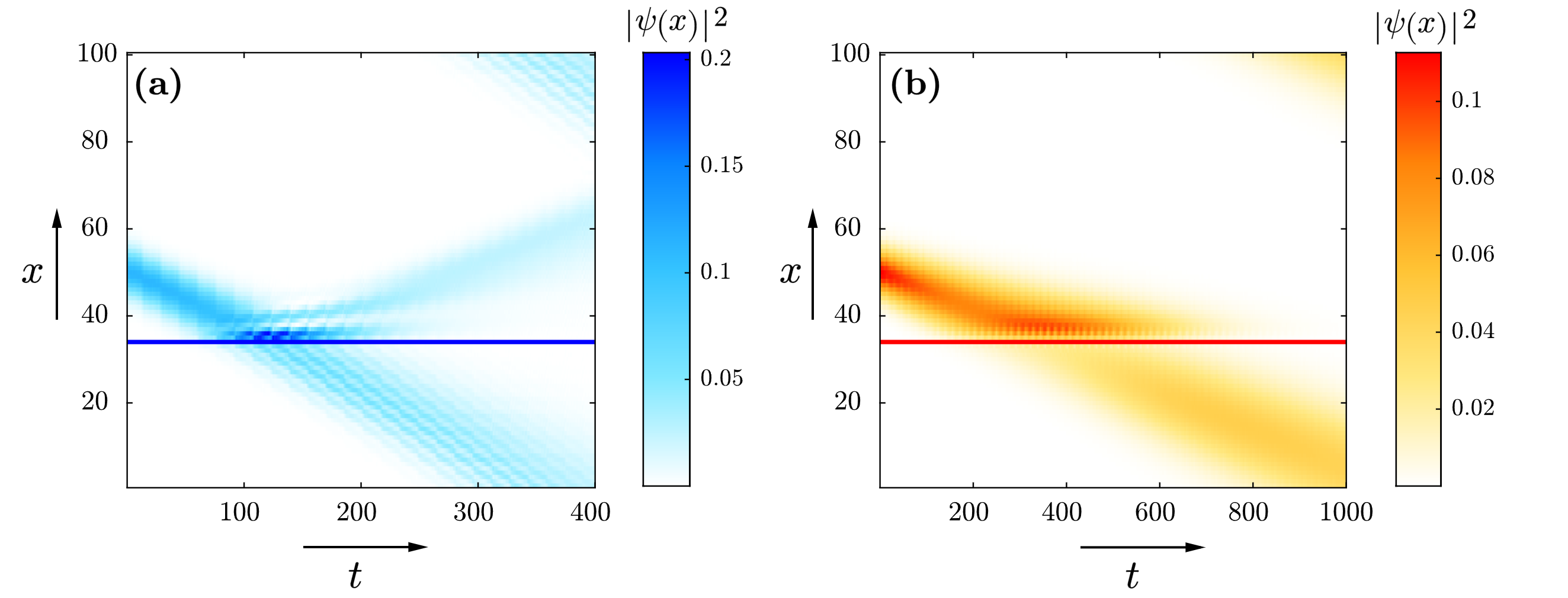}
    \caption{Hitting an impurity at $x_{\text{imp}}=L/3$  (Parameters: $q=3$, $\Omega=0.4,u=v=1,L=100$). 
 (a) Hermitian case. The initial wavepacket is Gaussian. After hitting the impurity, back-scattering is observed. (b) Non-Hermitian case where on-site dissipation $(\gamma_a=-1.2)$ is turned on. After hitting the impurity, the wavepacket passes through it without back-scattering.}
    \label{fig:robustness}
\end{figure}
\section{Wavepacket dynamics}
\subsection{Robustness of the one-way transport}
\label{ap:robust}
In the main text, we claim that the one-way transport in the Floquet AAH chain with dissipation is robust against impurities. 
It can be seen as the consequence of the Floquet emergent non-Hermitian skin effect, where the forward and backward propagation has different lifetimes. 
Here we will demonstrate this using wavepacket dynamics with a single impurity site placed in the system.

The impurity is simulated by a potential wall of magnitude $V_{\text{imp}}=u/10$ and placed at $x_{\text{imp}}=L/3$. 
The initial wavefunction is taken as a Gaussian wavepacket with the same parameters as in Fig.~\ref{fig:Fig1} of the main text. 
We first put the wavepacket into the Hermitian system ($\gamma_a=0$). 
After hitting the impurity line, a portion of the wavepacket is back-scattered by the impurity, and the rest crosses the potential wall (Fig.~\ref{fig:robustness}(a)).
This shows that a Hermitian Floquet system is not immune to impurities.
Now, we put the wavepacket into the non-Hermitian system with dissipation $(\gamma_a=-1.2)$. 
As demonstrated in Fig.~\ref{fig:robustness}, the back-scattering channel is suppressed. 
The impurity is basically 'invisible' to the wavepacket in the system with dissipation. From this single-impurity simulation, it can be anticipated that the propagation in the real system should also be immune to impurities, as the back-scattering channel has a shorter lifetime and quickly decays. Thus, we provide strong evidence for the robustness of the unidirectional transport.

\begin{figure}[h]
    \centering
    \includegraphics[width=\linewidth]{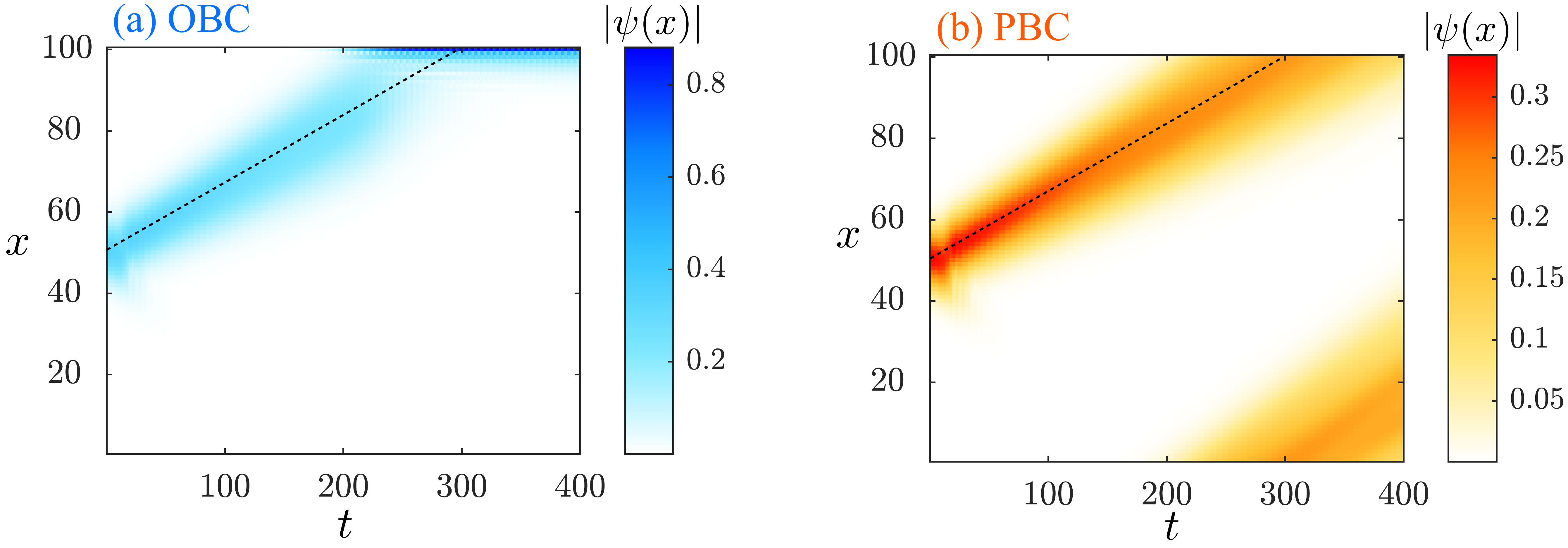}
    \caption{(a) Open boundary condition. Notably, the wavepacket will accumulate at the end of the system; (b) Periodic boundary condition; The bulk dynamics is independent of the boundary conditions.}
    \label{fig:rep_bc}
\end{figure}

Moreover, the evolution of the wavepacket in the bulk of the system is not affected by the boundary condition of the system.
We used periodic boundary conditions in our manuscript when doing wavepacket dynamics simulations.
If we switch to open boundary conditions, the wavepacket evolution in the bulk of the system will not be affected.
There will only be differences at the boundary of the system.
These are shown in Fig.~\ref{fig:rep_bc}.
Actually, it is proved rigorously that the propagator in the thermodynamic limit is independent of the boundary condition. [Mao et al., Phys. Rev. B 104, 125435].

\subsection{Dependence on the dissipation strength}

\begin{figure}[t]
    \centering
    \includegraphics[width=1\linewidth]{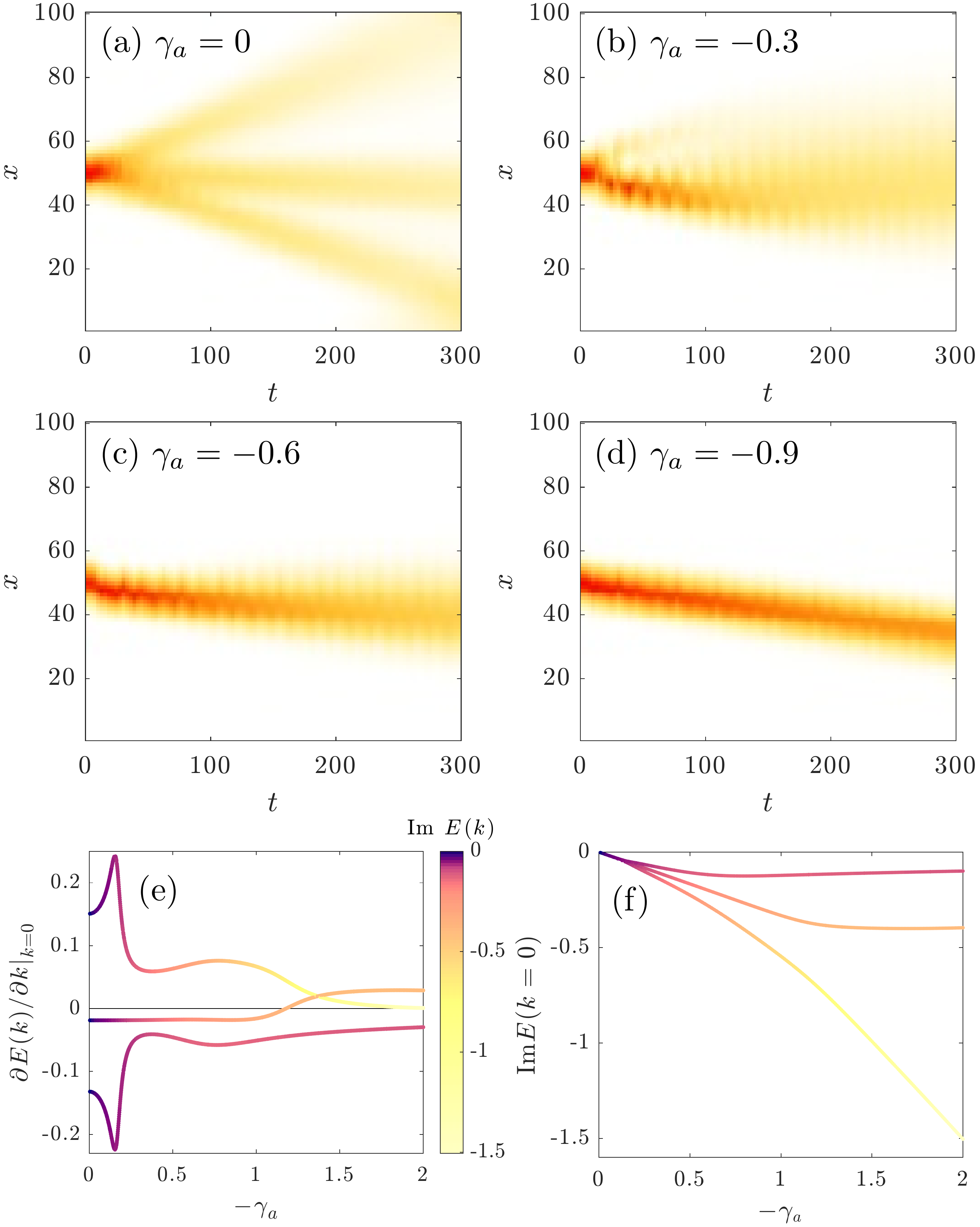}
    \caption{(a)-(d) The behavior of the wavepacket as we gradually increase the strength of dissipation $\gamma_a$. They all use the set of parameters as Fig.1 (c) in the main text and share the same initial condtion. In (a), there is no dissipation. The wavepacket acquires three velocities according to the slopes of the three Floquet bands at $k=0$. From (b)-(d), the propagation becomes one-way to the $-x$ direction, which agrees with the information provided by the GBZ; (e) The band velocities at $k=0$ as a function of dissipation strength $\gamma_a$; (f) The imaginary part of the three bands at $k=0$ as a function of the dissipation strength $\gamma_a$.}
    \label{fig:SF2}
\end{figure}

In this section, we demonstrate the behavior of the wavepacket dynamics as we gradually increase the dissipation strength. A Gaussian wavepacket is still chosen as the initial wavefunction. In the main text, the inner degree of freedom $\bfu$ is fined-tuned so that one of the three bands is excited. But if we take $\bfu$ randomly, then generically all three bands will be excited, causing the wavepacket to split into three parts with different velocities. (Fig.~\ref{fig:SF2}(a)) When the non-Hermitian parameter $\abs{\gamma_a}$ is gradually increased, the three eigenvalues will possess different imaginary parts as shown in Fig.~\ref{fig:SF2}(e), manifested as different lifetimes of the velocity components. As a result, the band with the largest imaginary part survives in the end as depicted in Fig.~\ref{fig:SF2}(a)-(d). In Fig.~\ref{fig:SF2}(f), the variation of the imaginary part with respect to the dissipation strength is depicted. After a critical strength at approximately $\gamma_c=-0.2$, the imaginary parts of the three bands split, and the band with the largest imaginary part dominates. Since we have calculated the stroboscopic generalized Brillouin Zone of our system which is totally included within the Brillouin Zone, the velocity of the band which dictates the dynamics must have the negative velocity (towards $-x$ direction), which totally agrees with our calculation in Fig.~\ref{fig:SF2}(e).

\subsection{Incommensurate AAH models}
\label{subsec:incom}
In the main text, only commensurate models are discussed where the flux $\phi$ is given by a rational number $p/q$. The incommensurate models are quasicrystals, and thus cannot be described by band theory. In this section, however, we show that the incommensurate model also possesses similar one-way transport properties. The idea is that any irrational number can be approximated by a rational number with arbitrarily high precision. In principle, the dynamic properties of the driven AAH model are not sensitive to the infinitesimal difference of $\phi$. Therefore, the wavepacket dynamics of an incommensurate model should, in principle, be similar to the commensurate model with a close value of $\phi$.

\begin{figure}[h]
    \centering
    \includegraphics[width=1\linewidth]{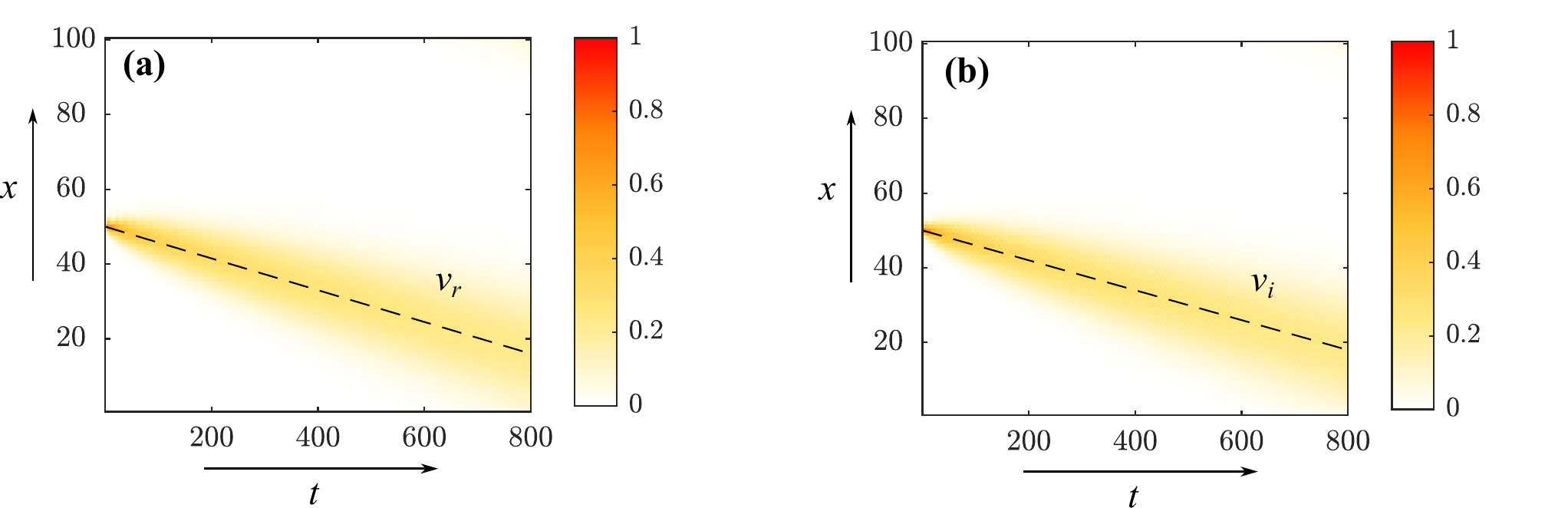}
    \caption{The wavepacket simulation for a commensurate model and a very close incommensurate model.}
    \label{fig:SF3}
\end{figure}
\begin{figure*}[t]
    \centering
    \includegraphics[width=0.77\linewidth]{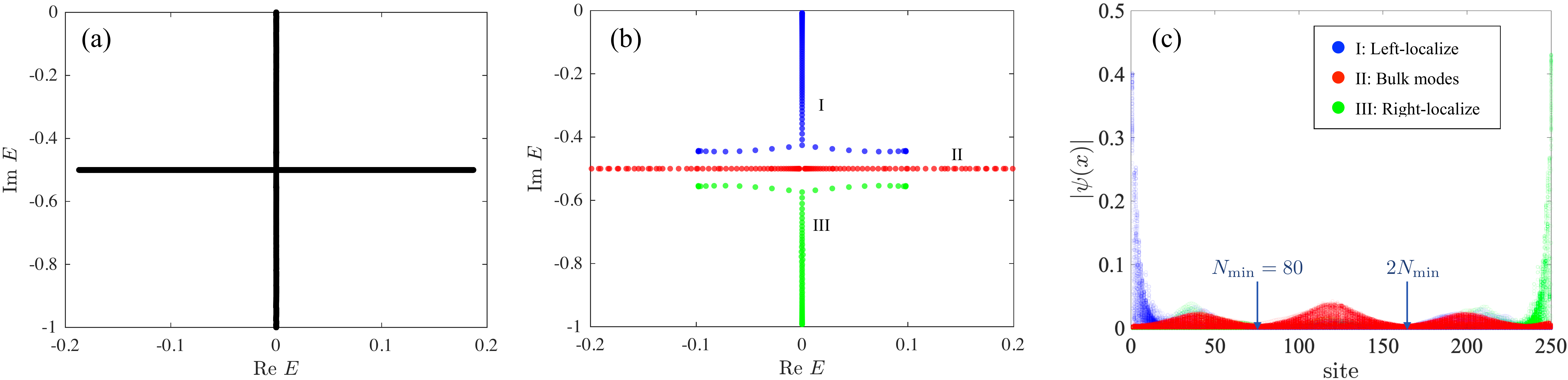}
    \caption{Emergence of skin mode in incommensurate models slightly deviates from $\phi=1/2$. (a) The spectrum of $\phi=1/2$; (b) The spectrum of $\phi=1/2-\epsilon$ and $\epsilon=\pi/1000$; (c) Corresponding eigenfunctions of bands I, II and III. Valleys around $N\simeq N_{\text{min}}$ can be seen.}
    \label{fig:incomm_spec_wav}
\end{figure*}
To show this, we take the same commensurate model as in the main text with rational flux $\phi_r=1/3$ and a very close incommensurate model with irrational $\phi_i$
\begin{equation}
    \phi_i=\frac{1}{3}\times\frac{\sqrt{2}}{1.415}\approx\frac{1}{3}-0.0002\approx\phi_r.
    \label{eq:incom}
\end{equation}

Then we perform the same wavepacket simulation for the $\phi_i$ and $\phi_r$ models. The initial wavefunction is chosen as the delta function. As shown in Fig.~\ref{fig:SF3}, when applying the same dissipation $\gamma_a=-1.2$, the dominant velocity components are $v_r=-0.0400$ and $v_i=-0.0425$ for the commensurate and the incommensurate models. The deviation $\sim 6\%$ is due to the difference of the $\phi_i$ and $\phi_r$ as given in (\ref{eq:incom}).
In experiments, there must be small deviations from the ideal rational $\phi$, making the real system incommensurate. However, we showed here that the dynamical behaviors would be almost identical to the commensurate model which has a similar $\phi$.

That being said, however, if the incommensurate period is very close to the reciprocal limit $q=2$, the reciprocity will be weakly broken.
In short, when $\phi$ slightly deviates from $1/2$, the constraint of reciprocal symmetry is relieved and skin modes will emerge. 
However, this effect can be treated as perturbation in the sense that the non-reciprocity of the skin modes is proportional to the deviation of $\phi$ from $1/2$, which means that in dynamics the initial wavepacket will gain a weak one-way drift velocity. 
In the following, we will elaborate on these statements and present our calculation of the incommensurate models around $\phi\simeq 1/2$. 
\\

\begin{figure}[h]
    \centering
    \includegraphics[width=\linewidth]{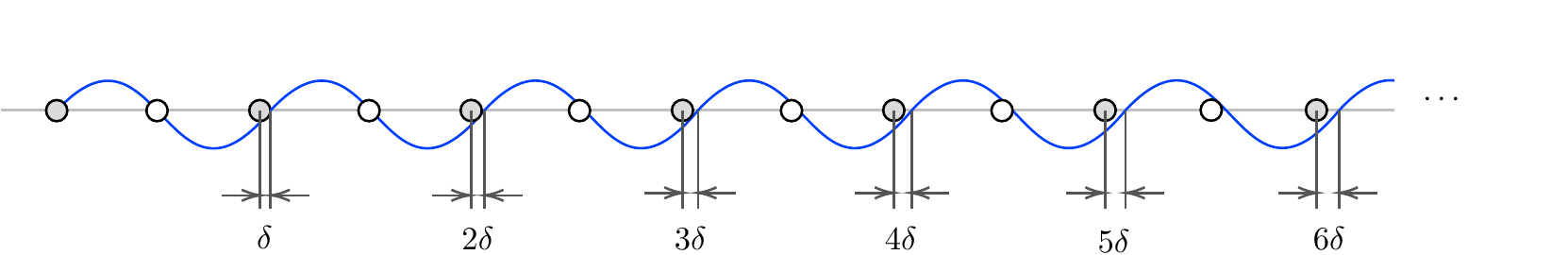}
    \caption{Sketch of incommensurate potential with period $\lambda=2+\delta$}
    \label{fig:sketch_incomm}
\end{figure}

 When $\phi = p/q =1/2$, the Floquet effective Hamiltonian follows the reciprocity condition
\begin{equation}\label{eq:reciprocity}
    H_F^T(k)=H_F(-k).
\end{equation}
However, when the period of the potential slightly deviates from the $\phi=1/2$, the reciprocity is weakly broken, and consequently, skin effects will emerge. 
As an example, consider an incommensurate model with an on-site potential of period
\begin{equation}
    \phi_i=\frac{1}{2+\delta} = \frac{1}{2}-\frac{\pi}{1000}\simeq 0.49686
\end{equation}
such that the period of the potential is $2+\delta\simeq 2.01265$. However, the period of the on-site dissipation is still $2$, as we introduce dissipation for the first site of every two sites. The model with the incommensurate potential is illustrated in Fig.~\ref{fig:sketch_incomm}. 

 At a large system size, a periodic interference pattern can be observed, as a result of stacking the potential and the lattice with incommensurate periods.
The wavelength of the interference pattern $N$ is determined by
\begin{equation}
    N\delta=1\ \Rightarrow\ N_{\text{min}}=[N]=80
    \label{eq:incomm_per}
\end{equation}
where $N_{\text{min}}$ is the closest integer to $N$.

Next, we calculate the spectrum of this incommensurate model and compare it with the $\phi=1/2$ case.
The spectra of the $\phi=1/2$ model under open and periodic boundary conditions coincide with each other, signifying the absence of the skin effect (Fig.~\ref{fig:incomm_spec_wav}). 
For $\phi=1/(2+\delta)\simeq1/2-\delta$, if $\delta$ is a small irrational number, then we can calculate the open boundary condition eigenvalues as shown in Fig.~\ref{fig:incomm_spec_wav} (b).
Notably, the open-boundary eigenvalues ramify into three continuum bands, labeled as I, II, and III. 
By plotting the corresponding eigenfunctions, we found that the eigenmodes in band I are left-localized, the eigenmodes in band II are right-localized, and the eigenmodes in band III are extensive in the bulk. 
Specifically, valleys can be observed from the distribution of wavefunctions at around every $N_{\text{min}}=80$ lattice sites, as shown in Fig.~\ref{fig:incomm_spec_wav}(c), which directly corresponds to the period of the incommensurate potential we calculated in Eq.~(\ref{eq:incomm_per}).
\\

 Similar results can be obtained for arbitrary $\phi= 1/2+\epsilon,\epsilon\neq 0$, i.e., as long as the reciprocity at $\phi=1/2$ is broken, the skin effect will show up. 
Moreover, the strength of the skin effect, quantified by the localization length and the splitting of the spectrum, is proportional to the deviation $\epsilon\equiv\phi-1/2$. 
This can be seen if we treat $\epsilon$ as a perturbation:
\begin{equation}
    \hat{H}(t)=\hat{H}_0(t)+\epsilon\hat{H}_1(t)
\end{equation}
\begin{figure*}[t]
    \centering
    \includegraphics[width=0.77\linewidth]{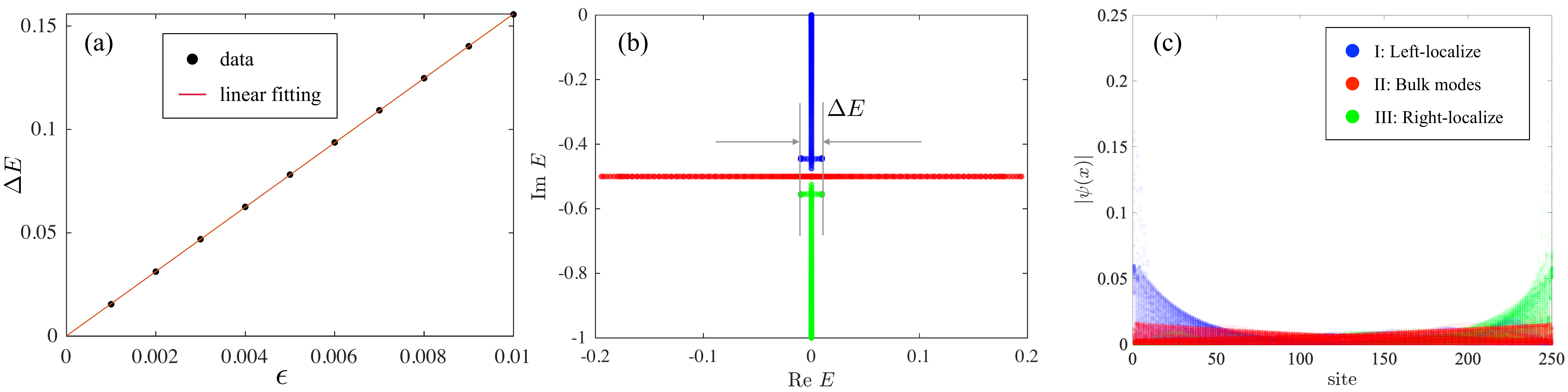}
    \caption{(a) The energy splitting $\Delta E$ as a function of the deviation $\epsilon$; (b) The spectrum of $\phi=1/2+\epsilon$ and $\epsilon=\pi/10000$; (c) Corresponding eigenwavefunctions.}
    \label{fig:small_dev}
\end{figure*}
Then, we can expand the effective Hamiltonian to the first order:
\begin{equation}
\begin{aligned}
     \hat{H}_F&=\frac{\mathrm{i}}{T}\log\left\{ \hat{\mathcal{T}}\exp\left[-\mathrm{i}\int_0^T \mathrm{d}\tau\left(\hat{H}_0(\tau)+\epsilon\hat{H}_1(\tau)\right)\right]\right\}\\
     &:=\hat{H}_{F,0}+\epsilon\hat{H}_{F,1}
\end{aligned}
\end{equation}
and the perturbation to the eigenenergies is given by
\begin{equation}
    \Delta E = \epsilon\mel{\psi}{\hat{H}_{F,1}}{\psi}\propto\epsilon.
\end{equation}

We numerically verified this by calculating the energy splitting $\Delta E$ of bands I and III as shown in Fig.~\ref{fig:small_dev} (b). 
As shown in Fig.~\ref{fig:small_dev} (a), the splitting $\Delta E$ is indeed proportional to the deviation $\epsilon$ when $\epsilon$ is small.
Furthermore, the value of $\Delta E$ is directly related to the localization length of the skin modes in bands I and III.
The localization length of the skin modes (colored blue and green) in Fig.~\ref{fig:small_dev} (c) is longer than that in Fig.~\ref{fig:incomm_spec_wav}, since a smaller deviation $\epsilon$ leads to a smaller energy splitting $\Delta E$.

\section{Numerical calculation of the Stroboscopic Generalized Brillouin Zone}\label{ap:FGBZ}

\begin{figure*}[t]
    \centering
    \includegraphics[width=\linewidth]{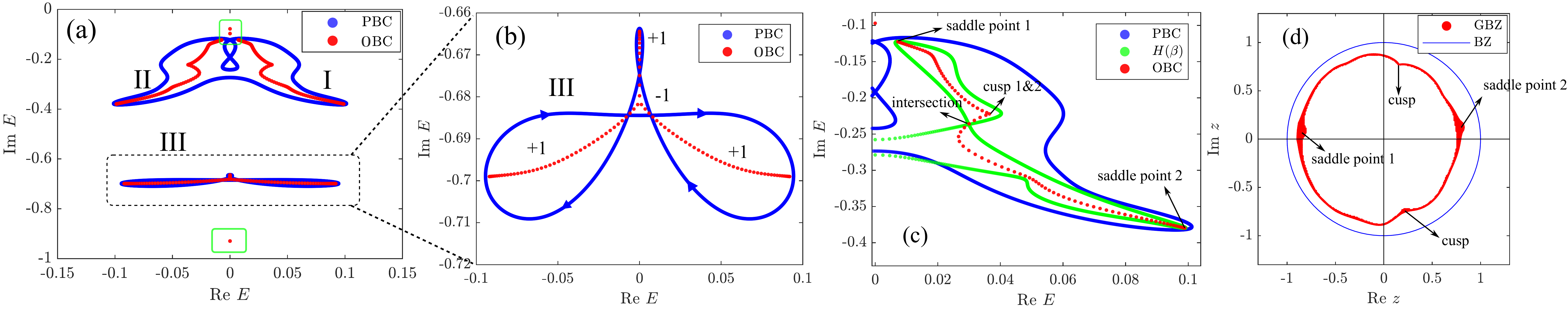}
    \caption{Key features of the stroboscopic GBZ: saddle points and cusps. (a) The PBC and OBC spectra, with topological edge states marked by the green boxes and the three bands are labeled by I, II, and III. (b) Local magnification of band III, and spectral winding numbers of each part. (c) Spectral features: Cusps, saddle points, and intersection with $H(\beta)$. (d) Numerically calculated Floquet GBZ, with corresponding saddle points and cusps.}
    \label{fig:AF10}
\end{figure*}

In this section, we provide details about the numerical calculation of the stroboscopic GBZ and present characteristic features of the calculated GBZ, such as saddle points and cusps. First, the complex energy spectrum of the effective Hamiltonian we studied in the main text is presented in Fig.~\ref{fig:AF10} (a). All complex eigenvalues under PBC and OBC are shown in (a), where the three non-Bloch bands are labeled as I, II, and III. Also, since the $q=3$ model is topologically nontrivial, there will be three topological edge states, boxed by the green squares. The isolated band III is magnified and shown in Fig.~\ref{fig:AF10} (b). The OBC eigenvalues of band III have nontrivial PBC spectral windings (has both $+1$ and $-1$ segments), indicating that there will be left and right localized skin modes \cite{zhang2020correspondence}. However, as there is a line gap between band III and the other two bands and the imaginary part of band III is the lowest, the spectral winding of band III has no effect on the long-time dynamics of our system. So in the following, we only plot the GBZ for Band I (Band II and Band I are symmetric about the imaginary axis, thus having the same GBZ). 

We analytically extend the momentum $k$ to the complex energy plane $k\to k+\rmi\kappa_c$ and take $k\in(-\pi,\pi)$, the trajectory of $\beta:=e^{i(k+\rmi\kappa_c)}$ will be a circle centered at the origin of the complex plane with radius $\rme^{-\kappa_c}$. Then we plot the eigenvalues of $\cH(\beta)$ where $\beta$ goes through this circle. Whenever the circle intersects with GBZ, the eigenvalues of $\cH(\beta)$ will also intersect with the OBC spectrum on the complex energy plane. 
Furthermore, since the OBC spectrum forms some arcs on the complex energy plane, every point on the OBC spectrum corresponds to two $\beta$ on GBZ. 
The OBC, PBC, and $\cH(\beta)$ spectra are plotted in Fig.~\ref{fig:AF10} (c) using different colors. 
The intersection between OBC and $\cH(\beta)$ as we mentioned can be observed. 
We scan the value of $\kappa_c$ as well as $k$ and search for all such intersections and plot them on the complex plane, which composes the stroboscopic GBZ. 
The numerical calculated GBZ has a finite broadening, as a result of the tolerance of the searching algorithm. 
Some essential features can also be read off from the obtained GBZ. 
First, the endpoints of the OBC spectrum must be saddle points on the GBZ~\cite{longhi2019probing}. 
They are captured by the numerical GBZ as shown in Fig.~\ref{fig:AF10}(d). 
Moreover, there is a cusp on the OBC spectrum. 
Since the OBC spectrum is a shrunken loop, this cusp corresponds to a pair of generalized momenta on the stroboscopic GBZ. 
They can also be observed in Fig.~\ref{fig:AF10}(d).

To conclude, the numerical method we adopted successfully captures the shape of the stroboscopic GBZ as the cusps and saddle points of the OBC spectrum can be reflected on the obtained GBZ. 
However, such searching algorithm is time-consuming and the detailed shape of the stroboscopic GBZ does not affect the long-time dynamics of the system. 
Practically, we only need to know the radial range of the stroboscopic GBZ to determine the propagation direction of the wavepacket. 
Therefore, we only need to scan $\kappa_c$ to see whether the triple degenerate intersection of $\cH(\beta)$ and OBC spectrum can be found, thus determining the radial range of the stroboscopic GBZ.
\section{Long time dynamical feature}
\label{ap:LongTime}
In the main text, we use the %Lyapunov exponent
time-average decay rate of the propagator as a dynamical probe of the emergent non-Hermitian skin effect. 
In this section, we show detailed proof that %the Lyapunov exponent 
this quantity can capture the saddle points of the OBC spectrum which happens to possess the largest imaginary part, thus dictating the long-time dynamics of our system. 
This quantity can be seen as the Floquet version of the zero-drift Lyapunov  that has been discussed in static band systems in Ref.\cite{longhi2019probing}.
In the main text, we defined the decay rate as
\begin{equation}
    G(x,t)=\mel{x_0+x}{\hat{\mathcal{T}}\rme^{-\rmi\int_0^t\hat{H}(\tau)\dd\tau}}{x_0},~\lambda=\overline{\partial_t\log\abs{G(x,t)}}
\end{equation}
At the stroboscopic level, we replace the time-dependent Hamiltonian with the Floquet effective Hamiltonian $\hat{H}_F$.
Then, we have
\begin{equation}\label{eq:Lya_evo}
\begin{aligned}
    G(x,t)&=\mel{x_0+x}{\rme^{-\rmi \hat{H}_{F} t}}{x_0}\\
    &=\int\dd k\int\dd k'\braket{x_0+x}{k}\mel{k}{\rme^{-\rmi \hat{H}_{F} t}}{k'}\braket{k'}{x_0}\\
    &=\int_{-\pi}^{\pi}\frac{\dd k}{2\pi}~\rme^{\rmi k x-\rmi E(k) t}\\
    &=\oint_{\text{BZ}}\frac{\dd z}{2\pi\rmi}z^{x-1}\rme^{-\rmi E(z)t}
\end{aligned}
\end{equation}
where in the last step we have replaced the variable $z=\rme^{\rmi k}$. 
To see the long-time dynamics, we can make use of the stationary phase approximation: (see, for instance, Chap.~1 of Ref.\cite{guillemin1990geometric})
\begin{equation}
    \oint_C\dd z~f(z)\rme^{\rmi g(z)}=\sum_s f(z_s)\rme^{\rmi g(z_s)}
\end{equation}
where $z_s$ stands for the saddle point satisfying $\partial_z \cH(z_s)=0$. 
However, this approximation should not be used directly, because generically there are no saddle points sitting on the BZ. 
But we do know that the GBZ does contain saddle points that correspond to the endpoints of the OBC spectrum \cite{longhi2019probing}. Notice that from equation (\ref{eq:Lya_evo}), the integrand contains no other singularities except for $z=0$. Hence we are allowed to deform the integration path from BZ to GBZ to capture the saddle points
\begin{equation}
    G(x,t)=\oint_{\text{GBZ}}\frac{\dd z}{2\pi\rmi}z^{x-1}\rme^{-\rmi E(z)t}\sim\rme^{-\rmi E(z_s)t}.
\end{equation}
After a long time, only the saddle points with the largest imaginary part dominate, so according to the definition of the  time-average decay rate, we get
\begin{equation}
    G(x,t)\sim\rme^{\Im E(z_s) t},~\lambda=\Im{E(z_s)}.
\end{equation}
\section{Directional control by changing frequency}

In this section, we show that the direction of propagation in our system can be easily controlled by changing the driving frequency.
In the experimental setting~\cite{cheng2020experimental}, the driving frequency of the system is proportional to the relative speed of the top and bottom layers of the acoustic waveguides.
Moreover, the PBC spectrum of the system can have dramatic differences at different driving frequencies, leading to different dominant velocities of wavepacket propagation.
Therefore, the propagation speed and direction can be easily manipulated simply by moving the top layer at different speeds.

\begin{figure}[t]
    \centering
    \includegraphics[width=\linewidth]{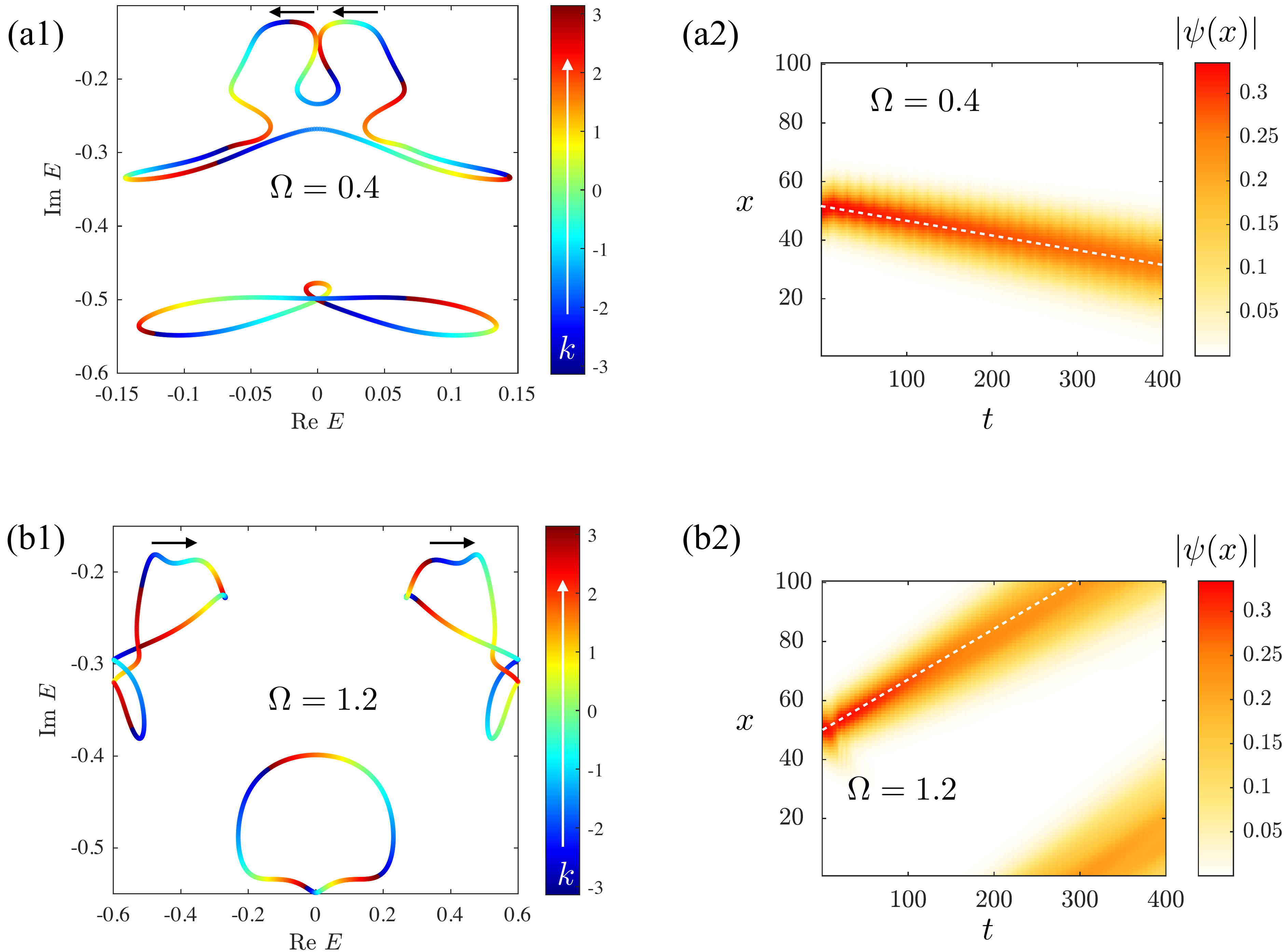}
    \caption{Controlling propagation direction by changing driving frequency. (a) When frequency $\Omega=0.4$, the PBC spectrum (a1) shows the propagation is to the $-x$ direction, confirmed by the wavepacket simulation in (a2); (b) When frequency $\Omega=1.2$, the PBC spectrum (b1) shows the propagation is to the $+x$ direction, confirmed by the wavepacket simulation in (b2) }
    \label{fig:r1}
\end{figure}

In Figs.~\ref{fig:r1} (a1) and (b1), the spectra of the effective Floquet Hamiltonian $H_F(k)$ under driving frequencies $\Omega=0.4$ and $\Omega=1.2$ are plotted.
The $k$-dependence of the spectrum is indicated by the colors.
The propagation direction under long-time evolution is determined by the parts of the spectrum with the largest imaginary part. 
The group velocities, as indicated by the black arrows, have opposite directions in these two systems at different driving frequencies. 
Numerical verification in Figs.~\ref{fig:r1} (a2) and (b2) confirms that the wavepacket indeed propagates in opposite directions. 
Notably, the perturbation of the driving frequency $\Omega$ does not cause abrupt changes in the quasienergy spectrum of $H_{F}(k)$, which implies that one-way propagation remains stable within a range of driving frequencies. 
It can be expected that there are potential applications for controlling the (robust) propagation direction by tuning the driving frequency. 

\bibliography{refs.bib}

%%%%%%%%%%%%%%%%%%%%%%%%%%%%%%%%%%%%%%%%%%%%%%%%%%%%%%%%%%%%%%%%%%%%%%%%%%%%
\onecolumngrid
\newpage
\renewcommand{\theequation}{S\arabic{equation}}
\renewcommand{\thefigure}{S\arabic{figure}}
\renewcommand{\thetable}{S\arabic{table}}
\setcounter{equation}{0}
\setcounter{figure}{0}
\setcounter{table}{0}

%%%%%%%%%%%%%%%%%%%%%%%%%%%%%%%%%%%%%%%%%%%%%%%%%%%%%%%%%%%%%%%%%%%%%%%%%%%%
\end{document}